\documentclass[11pt]{amsart}
\usepackage{latexsym,amssymb,amsmath,amscd,amsthm}
\numberwithin{equation}{section}
\theoremstyle{remark}

\newcommand{\bq}{\begin{equation}}
\newcommand{\bea}{\begin{array}}
\newcommand{\eea}{\end{array}}

\newcommand{\ga}{\alpha}
\newcommand{\gep}{\epsilon}
\newcommand{\gD}{\Delta}
\newcommand{\gl}{\lambda}

\newcommand{\gb}{\beta}

\newcommand{\mf}{\mathfrak}
\newcommand{\mc}{\mathcal}

\newcommand{\gG}{\Gamma}

\newcommand{\gag}{\gamma}
\newcommand{\gd}{\delta}
\newcommand{\pp}{\partial}

\newcommand{\tl}{\tilde}
\newcommand{\na}{\nabla}

\newcommand{\bs}{\blacksquare}

\newcommand{\bgs}{\bigstar}

\newcommand{{\DDD}}{D\!\!\!\!\!\!-}

\title{ON STABILITY, FLUCTUATIONS, AND QUANTUM MECHANICS }

\author{Robert Carroll\\University of Illinois, Urbana, IL 61801}

\date{August, 2008\thanks{email: rcarroll@math.uiuc.edu}}

\begin{document}

\bibliographystyle{plain}

\begin{abstract} 
We review an important stability approach to quantization by Rusov and Vlasenko and
indicate possible comparison of fluctuations to standard situations involving a quantum potential.
\end{abstract}

\maketitle
\tableofcontents


\section{INTRODUCTION}
\renewcommand{\theequation}{1.\arabic{equation}}
\setcounter{equation}{0}

In \cite{rsov,rsvo} (which are the same modulo typos and conclusions) one sketches how the work
of Chetaev \cite{chet,chte,chee} (based in particular on classical results of Poincar\'e \cite{poin}
and Lyapunov \cite{lyap}) might allow one to relate stability of classical Hamiltonian systems
to quantum mechanics.
We review here some of the
arguments 
(cf. also \cite{ahmd,ahun,bypt,ghad,ghsn,rumy}
for generalities on the Poincar\'e-Chetaev equations).
\\[3mm]\indent
One recalls that holonomic systems involve an agreement of the degrees of freedom
with the number of independent variables (cf. \cite{whit}).  Then following \cite{chet}
consider a holonomic system with Hamiltonian coordinates
\bq\label{1.1}
\frac{dq_j}{dt}=\frac{\pp H}{\pp p_j};\,\,\frac{dp_j}{dt}=-\frac{\pp H}{\pp q_j}
\end{equation}
and think of perturbations $({\bf 1A})\,\,q_j=q_j(t)+\xi_j$ and $p_j=p_j(t)+\eta_j$.  Denoting then
$q_j\sim q_j(t)$ and $p_j\sim p_j(t)$ one has
\bq\label{1.2}
\frac{d(q_j+\xi_j)}{dt}=\frac{\pp H(t,q_i+\xi_i,p_i+\eta_i)}{\pp p_j};\,\,\frac{d(p_j+\eta_j)}{dt}
=-\frac{\pp H(t,q_i+\xi_i,p_i+\eta_i)}{\pp q_j}
\end{equation}
(note that no connection is made a priori to a Schr\"odinger equation - SE).
Expanding and using (1.1) gives
\bq\label{1.3}
\frac{d\xi_j}{dt}=\sum\left(\frac{\pp^2H}{\pp p_j\pp q_i}\xi_i+\frac{\pp^2H}{\pp p_j\pp p_i}\eta_i
\right)+X_j;
\end{equation}
$$\frac{d\eta_j}{dt}=-\sum\left(\frac{\pp^2H}{\pp q_j\pp q_i}\xi_i+\frac{\pp^2H}{\pp q_j\pp p_i}\right)\eta_i+Y_j$$
where the $X_j,Y_j$ are higher order terms in $\xi,\,\eta$.  The first approximations (with
$X_j=Y_j=0$) are referred to as Poincar\'e variational equations.  Now given 
stability questions relative to functions $Q_s$
of $(t,q,p)$ one writes
\bq\label{1.4}
x_s=Q_s(t,q_i+\xi_i,p_i+\eta_i)-Q_s(t,q_i,p_i)=\sum \left(\frac{\pp Q_s}{\pp q_i}\xi_i+\frac
{\pp Q_s}{\pp p_i}\eta_i\right)+\cdots
\end{equation}
which implies
\bq\label{1.5}
\frac{dx_s}{dt}=\sum\left(\frac{\pp Q_s'}{\pp q_i}\xi_i+\frac{\pp Q_s'}{\pp p_i}\eta_i\right)+\cdots
\end{equation}
where
\bq\label{1.6}
Q_s'=\frac{\pp Q_s}{\pp t}+\sum\left(\frac{\pp Q_s}{\pp q_i}\frac{\pp H}{\pp p_i}-\frac{\pp Q_s}{\pp p_i}
\frac{\pp H}{\pp q_i}\right)
\end{equation}
Given $1\leq s\leq 2k$ and $1\leq i,j\leq k$ one can express the $\xi_i,\,\eta_i$ in terms of $x_s$
and write $({\bf 1B})\,\,(dx_s/dt)=X_s$ (normal form) with $X_s(0)=0$.
For equations ({\bf 1B}) with $1\leq s\leq n$, for sufficiently small perturbations 
$\gep_j,\,\gep'_j$ one assumes there exists some system of initial values $x_{s^0}$ with
$\sum x^2_{s^0}<A$ for an arbitrarily small A (with perturbations $\gep_j,\,\gep_j'\leq E_j,\,E_j'$).
Further for arbitrarily small $E_j,\,E_j'$ one assumes it is possible to find A as above such that
there exists one or more values $\gep_j,\,\gep_j'$ with absolute values $\leq E_j,\,E_j'$.
Under these conditions the initial values of $x_s$ play the same role for stability as the
$\gep_j,\,\gep_j'$ and one assumes this to hold.  One assumes also convergent power series
for the $X_s$ etc.  Then Lyapunov stability means that for arbitrary small A there exists $\gl$ such that for all perburbations $x_{s^0}$ satisfying $\sum x^2_{s^0}\leq \gl$ and for all $t\geq t_0$
one has $\sum x_s^2<A$ (i.e. the unperturbed motion is stable).
Next one considers $t\geq t_0$ and $\sum x_s^2\leq H$ and looks for a sign definite 
(Lyapunov) function V (with $V'=\pp_tV+\sum_1^nX_j(\pp V/\pp x_j)$ then sign definite of opposite sign or zero).  If such a function exists the unperturbed motion is stable (see \cite{chet} for proof).
\\[3mm]\indent
We pick up the story now in \cite{chte} where relations between optics and mechanics are
illuminated.  Take a holonomic mechanical system with coordinates $q_i$ and conjugate
momenta $p_i$ with n degrees of freedom.  Assume the holonomic constraints are independent
of time and the forces acting on the system are represented by a potential function $U(q_i)$.
Let $({\bf 1C})\,\,T=(1/2)\sum_{i,j}g_{ij}p_ip_j$ denote the kinetic energy where the $g_{ij}=g_{ji}$
are not dependent explicitly on time.  Hamilton's equations have the form
\bq\label{1.7}
2T=\sum g_{ij}\frac{\pp S}{\pp q_i}\frac{\pp S}{\pp q_j}=2(U+E)
\end{equation}
where E represents a kinetic energy constant (the sign of U is changed in Section 2). 
Here the integral of (1.7) is $({\bf 1D})\,\,S(q_i,\ga_i)+c$ with
the $\ga_i$ constants and $({\bf 1E})\,\,||\pp^2S/\pp q_i\pp\ga_j||\ne 0$ while $({\bf 1F})\,\,
E=E(\ga_i)$.  According to the Hamilton-Jacobi theory the general solution of the motion equations
is given via $({\bf 1G})\,\,p_i=\pp S/\pp q_i$ and $\gb_i=-t(\pp E/\pp \ga_i)+\pp S/\pp\ga_i$ where
the $\gb_i$ are constants.  In order to determine a stable solution one looks at the Poincar\'e
variations
\bq\label{1.8}
\frac{d\xi_i}{dt}=\sum_j\left(\frac{\pp^2H}{\pp q_j\pp p_i}\xi_j+\frac{\pp^2H}{\pp p_j\pp p_i}\eta_j\right);
\end{equation}
$$\,\,\frac{d\eta_i}{dt}=-\sum\left(\frac{\pp^2H}{\pp q_j\pp q_i}\xi_j+\frac{\pp^2H}{\pp p_j\pp q_i}\eta_i
\right)$$
where H should be defined here via $({\bf 1H})\,\,H=T-U$.  For a stable unperturbed motion the differential equations for Poincar\'e
variations (1.8) must be reducible by nonsingular transformation to a system of linear
differential equations with constant coefficients all of whose characteristic values must be zero
(recall that the Lyapunov characteristic value $X[f]$ of $f$ is $X[f]=-\overline{lim}_{t\to\infty}[log
(|f(t)|)/t$ - cf. \cite{lyap,makl}).
In such perturbed motion, because of ({\bf 1G}) one has (recall $p_i\sim\pp S/\pp q_i$)
\bq\label{1.9}
\eta_i=\sum_j\frac{\pp^2S}{\pp q_i\pp q_j}\xi_j\,\,(i=1,\cdots,n)
\end{equation}
Hence 
\bq\label{1.10}
\frac{d\xi_i}{dt}=\sum_{j,s}\xi_s\frac{\pp}{\pp q_s}\left(g_{ij}\frac{\pp S}{\pp q_j}\right)\,\,
(i=1,\cdots,n)
\end{equation}
Note here that (1.8) involves $\sum g_{ij}p_ip_j-U$ so 
$$(\bgs)\,\,\frac{\pp H}{\pp p_i}=\sum g_{ij}p_j;\,\,\frac{\pp H}{\pp q_j}=\sum\frac{\pp g_{ij}}{\pp q_j}p_ip_j-\frac{\pp U}{\pp q_j}$$
and (1.10)  says
$$(\bgs\bgs)\,\,\frac{d\xi_i}{dt}=\sum \xi_s\left(\frac{\pp g_{ij}}{\pp q_s}\frac{\pp S}{\pp q_j}
+g_{ij}\frac{\pp^2S}{\pp q_s\pp q_j}\right)=$$
$$=\sum \xi_s\frac{\pp g_{ij}}{\pp q_s}\frac{\pp S}{\pp q_j}+\sum g_{ij}\eta_j$$
The second term here is $[\pp^2H/\pp p_i\pp p_j]\eta_j$ and we want to identify the term $\xi_s
(\pp g_{ij}/\pp q_s)(\pp S/\pp q_j)$ with $\pp^2H/\pp q_s\pp p_i\xi_s$.  However we can see that 
$\pp U/\pp p_i=0$ so $\xi_s(\pp^2H/\pp q_s\pp p_i)=\xi_s(2\pp^2T/\pp q_s\pp p_i)=\xi_s(\pp g_{ij}/\pp q_s)p_j$ 
confirming (1.10).
Here the $q_i,\,\ga_i$ are represented by their values in an unperturbed motion.
Now for a stable unperturbed motion let (1.10) be reducible by a nonsingular linear transformation
$({\bf 1I})\,\,x_i=\sum\gag_{ij}\xi_j$ with a constant determinant $\gG=||\gag_{ij}||$.  If
$\xi_{ir}\,\,(r=1,\cdots,n)$ are a normal system of independent solutions of (1.10) then
$({\bf 1J})\,\,x_{ir}=\sum_j\gag_{ij}\xi_{jr}$ will be the solution for the reduced system.  For a
stable unperturbed motion all the characteristic values of the solutions $x_{ir}\,\,(i=1,\cdots,n)$
are zero and consequently
\bq\label{1.11}
||x_{sr}||=C^*=||\gag_{sj}|| ||\xi_{jr}||=\gG Cexp\left[\int \sum\frac{\pp}{\pp q_i}\left(g_{ij}
\frac{\pp S}{\pp q_j}\right)dt\right]
\end{equation}
Consequently for a stable perturbed motion (cf. \cite{chet,lyap,makl})
\bq\label{1.12}
\sum\frac{\pp}{\pp q_i}\left(g_{ij}\frac{\pp S}{\pp q_i}\right)=0
\end{equation}
Finally one considers a solution $({\bf 1K})\,\,\Phi(-Et+S)$ of the HJ equation and for a stable
unperturbed solution, because of (1.12), (1.7), and ({\bf 1G}), one has
\bq\label{1.13}
\sum\frac{\pp}{\pp q_i}\left(g_{ij}\frac{\pp\Phi}{\pp q_j}\right)=\Phi'\sum\frac{\pp}{\pp q_i}\left(g_{ij}
\frac{\pp S}{\pp q_j}\right)
+\Phi''\sum g_{ij}\frac{\pp S}{\pp q_i}\frac{\pp S}{\pp q_j}=\frac{2(U+E)}{E^2}\frac{\pp^2\Phi}
{\pp t^2}
\end{equation}
which is a wave equation
\bq\label{1.14}
\frac{2(U+E)}{2E^2}\frac{\pp^2\Phi}{\pp t^2}=\sum\frac{\pp}{\pp q_i}\left(g_{ij}\frac{\pp\Phi}{\pp q_j}
\right)
\end{equation}
This indicates the analogy between Cauchy's theory of light and stable motions of 
holonomic conservative systems (cf. \cite{chet,chte,chee}).

\section{STABILITY APPROACH}
\renewcommand{\theequation}{2.\arabic{equation}}
\setcounter{equation}{0}

Following Rusov and Vlasenko one writes an integral of the
Hamilton-Jacobi (HJ) equation in the form $({\bf 2A})\,\,S=f(t,q_i,\ga_i)+A\,\,(i=1,\cdots,n)$ with
the $\ga_i$ arbitrary constants.  The general solution is then $({\bf 2B})\,\,p_i=\pp S/\pp q_i$ with
$\gb_i=\pp S/\pp \ga_i$ where the $\gb_i$ are new constants of integration.  The canonical equations of
motion are $dq_i/dt=\pp H/\pp p_i$ and $dp_i/dt=-\pp H/\pp q_i$ where H is the Hamiltonian
and under perturbations of the $\ga_i,\,\,\gb_i$ one writes $\xi_i=\gd q_i=q_i-q_i(t)$ and
$\eta_i=\gd p_i=p_i-p_i(t)$ and derives equations of first approximation
\bq\label{2.1}
\frac{d\xi_i}{dt}=\sum\frac{\pp^2H}{\pp q_j\pp p_i}\xi_j+\sum\frac{\pp^2H}{\pp p_j\pp p_i}\eta_j
\end{equation}
$$\frac{d\eta_i}{dt}=-\sum\frac{\pp^2H}{\pp q_j\pp q_i}\xi_j-\sum\frac{\pp^2H}{\pp p_j\pp q_i}\eta_j$$
as in (1.8).
By differentiating in $t$ one obtains then $({\bf 2C})\,\,C=\sum (\xi_s\eta'_s-\eta_s\xi'_s)$ where
C is a constant.  Also for given $\xi_s,\,\eta_s$ there is always at least one solution $\xi'_s,\,\eta'_x$
for which $C\ne 0$.  Stability considerations (as in (1.1))
then lead via $(\bgs)\,\,\eta_i=\sum (\pp^2S/\pp q_i\pp q_j)\xi_j$ and $({\bf 2D})\,\,H=(1/2)
\sum g_{ij}p_ip_j+U=T+U$ to 
\bq\label{2.2}
\frac{d\xi_i}{dt}=\sum \xi_s\frac{\pp}{\pp q_s}\left(g_{ij}\frac{\pp S}{\pp q_j}\right)
\end{equation}
(note in Section 1 $H\sim T-U$ following \cite{chte} but we take now $U\to -U$ to agree with
\cite{rsov,rsvo} - the sign of U is not important here).
According to \cite{rsov,rsvo}, based on results of Chetaev \cite{chte} (as portrayed in 
Section 1), it results that
$L=\sum(\pp/\pp q_i)[g_{ij}(\pp S/\pp q_j)]=0$ (as in (1.12)) for stability (we mention e.g.
\cite{chet,chte,chee,lyap,makl,nmst} for stability theory, Lyapunov
exponents, and all that). One also notes in \cite{rsov,rsvo} that a similar result occurs for $U\to U^*=U+Q$ for natural Q and it is assumed that it is Q which generates perturbations $\gd q,\,\gd p$.
\\[3mm]\indent
Now one introduces (in an ad hoc manner) a function $({\bf 2E})\,\,\psi=Aexp(ikS)$ in (1.12) where $k$ is constant and A is a real function of the coordinates $q_i$ only.  There results 
\bq\label{2.3}
\frac{\pp S}{\pp q_j}=\frac{1}{ik}\left(\frac{1}{\psi}\frac{\pp\psi}{\pp q_j}-\frac{1}{A}\frac{\pp A}{\pp q_j}\right)
\end{equation}
so that (1.12) becomes
\bq\label{2.4}
\sum_{i,j}\frac{\pp}{\pp q_i}\left[g_{ij}\left(\frac{1}{\psi}\frac{\pp\psi}{\pp q_j}-\frac{1}{A}
\frac{\pp A}{\pp q_i}\right)\right]=0
\end{equation}
On the other hand for the perturbed motion (with $U\to U^*=U+Q$) the HJ equation can be written
in the form
\bq\label{2.5}
\frac{1}{2k^2}\sum_{i,j}g_{ij}\left[\frac{1}{\psi}\frac{\pp\psi}{\pp q_i}-\frac{1}{A}\frac{\pp A}{\pp q_i}
\right]\left[\frac{1}{\psi}\frac{\pp\psi}{\pp q_j}-\frac{1}{A}\frac{\pp A}{\pp q_j}\right]=
\pp_tS+U+Q
\end{equation}
with $\pp_tS$ obtained via ({\bf 2E}).  Adding (2.4) and (2.5) yields
\bq\label{2.6}
\frac{1}{2k^2\psi}\sum_{i,j}\frac{\pp}{\pp q_i}\left(g_{ij}\frac{\pp\psi}{\pp q_j}\right)-\frac{1}{2k^2A}
\sum_{i,j}\frac{\pp}{\pp q_i}\left(g_{ij}\frac{\pp A}{\pp q_j}\right)-
\end{equation}
$$-\frac{1}{k^2A}\sum_{i,j}g_{ij}\frac{\pp A}{\pp q_j}\left(\frac{1}{\psi}\frac{\pp\psi}{\pp q_i}
-\frac{1}{A}\frac{\pp A}{\pp q_i}\right)-\frac{1}{ikA\psi}[A\pp_t\psi-\psi\pp_tA]-U-Q=0$$
as a necessary stability condition (in the first approximation).  Note (2.6) will not contain Q if A is
defined via
\bq\label{2.7}
\frac{1}{2k^2A}\sum_{i,j}\frac{\pp}{\pp q_i}\left(g_{ij}\frac{\pp A}{\pp q_j}\right)+
\frac{i}{kA}\sum_{i,j}g_{ij}\frac{\pp A}{\pp q_j}\frac{\pp S}{\pp q_i}-\frac{1}{ikA}\pp_tA+Q=0
\end{equation}
which means
\bq\label{2.8}
Q=-\frac{1}{2k^2A}\sum_{i,j}\frac{\pp}{\pp q_i}\left(g_{ij}\frac{\pp A}{\pp q_j}\right);\,\,
\pp_tA=-\sum_{i,j}g_{ij}\frac{\pp A}{\pp q_j}\frac{\pp S}{\pp q_i}
\end{equation}
A discussion of the physical content of (2.8) appears in \cite{rsov,rsvo} and given (2.8) the
stability condition (2.6) leads to
\bq\label{2.9}
\frac{i}{k}\pp_t\psi=-\frac{1}{2k^2}\sum_{i,j}\frac{\pp}{\pp q_i}\left(g_{ij}\frac{\pp\psi}{\pp q_j}\right)
+U\psi
\end{equation}
which is of course a SE for $k=1/\hbar$ (this is the place where quantum mechanics 
somewhat abruptly enters
the picture - see Remark 2.1).  
In fact for kinetic energy $({\bf 2F})\,\,T=(1/2m)
[p_1^2+p_2^2+p_3^2]$ (2.9) leads to 
\bq\label{2.10}
Q=-\frac{\hbar^2}{2m}\frac{\gD A}{A};\,\,\pp_tA=-\frac{1}{m}\sum\frac{\pp A}{\pp x_j}p_j;\,\,k=\frac{1}{\hbar}
\end{equation}
and (2.9) becomes
\bq\label{2.11}
i\hbar\pp_t\psi=-\frac{\hbar^2}{2m}\gD\psi+U\psi
\end{equation}
Going backwards now put the wave function $\psi=Aexp(iS/\hbar)$ in (2.11) to obtain 
via (1.12) and (2.8) the Bohmian equations
\bq\label{2.12}
\pp_tA=-\frac{1}{2m}[A\gD S+2\na A\cdot\na S]=-\na A\cdot\frac{\na S}{m};\,\,\pp_tS=
-\left[\frac{(\na S)^2}{2m}+U-\frac{\hbar^2}{2m}\frac{\na A}{A}\right]
\end{equation}
where the quantum potential QP is naturally identified.
\\[3mm]\indent
If one writes now $P=\psi\psi^*=A^2$ then (2.12) can be rewritten in a familiar form
\bq\label{2.13}
\pp_tP=-\na P\cdot\frac{\na S}{m};\,\,\pp_tS+\frac{(\na S)^2}{2m}+U-\frac{\hbar^2}{4m}
\left[\frac{\gD P}{P}-\frac{1}{2}\frac{(\na P)^2}{P^2}\right]=0
\end{equation}
That P is indeed a probability density is substantiated via a (somewhat vague) least action of perturbation principle of Chetaev \cite{chee} which takes the form
$({\bf 2G})\,\,\int Q|\psi^2|dV=min$ where $dV$ is a volume
element for the phase space ($\int |\psi|^2dV=1$) and this condition involves absolute stability
(one assumes that the influence of perturbative forces generated by Q is proportional to the
density $|\psi|^2=A^2$).
Write now, using ({\bf 2D}) 
\bq\label{2.14}
Q=-\pp_tS-U-T=-\pp_tS-U-\frac{1}{2}\sum g_{ij}\frac{\pp S}{\pp q_i}\frac{\pp S}{\pp q_j}
\end{equation}
Then if ({\bf 2E}) holds one can show that
\bq\label{2.15}
\frac{1}{2}\sum g_{ij}\frac{\pp S}{\pp q_i}\frac{\pp S}{\pp q_j}=-\frac{1}{2k^2\psi^2}\sum g_{ij}
\frac{\pp\psi}{\pp q_i}\frac{\pp\psi}{\pp q_j}+
\end{equation}
$$+\frac{1}{2k^2A^2}\sum g_{ij}\frac{\pp A}{\pp q_i}
\frac{\pp A}{\pp q_j}+\frac{ik}{2k^2A^2}\sum g_{ij}\frac{\pp A}{\pp q_i}\frac{\pp S}{\pp q_j}$$
Then for the first term on the right side substitute its value from the first stability condition
(2.4), then insert this relation into (2.15) and put the result into the equation (2.14) corresponding
to the variational principle; the result is then (2.6) and consequently the resulting structure 
expression and the necessary condition for stability coincide with (2.8) and (2.9).  This leads
one to conclude classical mechanics and the quantization (stability) conditions
represent two complementary procedures for description of stable motions of a physical system
in a potential field.  The authors cite an impressive list of references related to experimental
work supporting these kinds of conclusion.
\\[3mm]\indent
{\bf REMARK 2.1.}
The arguments in \cite{rsov,rsvo}
have seemed to be independent of the nature of the perturbations
beyond the important relation (1.9).  However the emergence of Q as a quantum potential
provides $2\na A/A=\na P/P\sim \gd p$ as a ``standard" momentum fluctuation.  
This seems to suggest some equivalence to standard perturbative models with a quantum 
potential (cf. \cite{c3,carr}) and perhaps forecasts the uncertainty principle in some sense
(see below).  The technique could perhaps provide an alternative approach to some results of \cite{gssn,gslf}
for example involving the generation of the SE from Hamiltonian theory via metaplectic coverings and short time propagators, etc.  The results reviewed here seem however to be perhaps too general 
although very attractive.
$\hfill\bs$

\section{THE QUANTUM POTENTIAL}
\renewcommand{\theequation}{3.\arabic{equation}}
\setcounter{equation}{0}

From Sections 1-2 we have the suggestion that given a stable Hamiltonian system with perturbations $\gd q$ and $\gd p$ generated by a ``potential" $Q\sim \gd U$ it follows that there is a Schr\"odinger equation (SE) with Q as the quantum potential (QP) which describes the motion.  
It seems therefore
appropriate to examine this in the light of other manifestations of the QP as in e.g.
\cite{c3,carr,c6,c7,crow,f4,f5,garb,gabr,gros,hall,hkr,hlal,hily,kani,kasc,regi}.
We note that following \cite{carr} one can reverse some arguments
involving the exact uncertainty principle (cf. \cite{c3,hall,hkr,hlal,regi}) to show that any SE 
described by a QP based on $|\psi|^2=P$ can be modeled on a quantum model of a classical
Hamiltonian H perturbed by a term $H_Q$ based on Fisher information, namely
\bq\label{3.1}
H_Q=\frac{c}{2m}\int\frac{(\na P)^2}{P}dx=\frac{c}{2m}\int P(\gd p)^2
\end{equation}
where $\gd p=\na P/P$.  This does not of course deny the presence of ``related" $x\sim q$
oscillations $\gd x\sim \gd q$ and in fact in Olavo \cite{olav} (cf. also \cite{c3}) Gaussian fluctuations
in $\gd q$ are indicated and related to $\gd p$ via an exact uncertainty relation $({\bf 3A})\,\,
(\overline{(\gd p)^2\cdot(\gd q)^2}=\hbar^2/4$.  We note that the arguments establishing 
exact uncertainty stipulate that the position uncertainty must be entirely characterized by 
$P=|\psi|^2$ (cf. \cite{c3,hall,hkr,hlal,regi}).
\\[3mm]\indent
{\bf REMARK 3.1.}
We recall here \cite{hest} (cf. also \cite{rrrl}) were it is shown that quantum mechanics can be considered
as a classical theory in which a Riemannian geometry is provided with the distance between
states defined with natural units determined via Planck's constant (which is the inverse of the
scalar curvature).$\hfill\bs$
\\[3mm]\indent
In \cite{bren} one shows that non-relativistic quantum mechanics for a free particle emerges
from classical mechanics via an invariance principle under transformations that preserve the
Heisenberg inequality.  The invariance imposes a change in the laws of classical mechanics
corresponding to the classical to quantum transition.  Some similarities to the Nottale
theory of scale relativity in a fractal space-time are also indicated (cf. \cite{c3,crnt,ntcl,nttl}).
There are relations here to the Hall-Reginatto treatment which postulates that the 
non-classical momentum fluctuations are entirely determined by the position
probability (as mentioned above).  In Brenig's work one derives this from an invariance principle
under scale transformations affecting the position and momentum uncertainties and preserving
the Heisenberg inequality.  One modifies the classical definition of momentum uncertainty
in order to satisfy the imposed transformation rules and this modification is also constrained
by conditions of causality and additivity of kinetic energy used by Hall-Reginatto.  This leads
to a complete specification of the functional dependance of the supplementary term corresponding
to the modification which turns out to be proportional to the quantum potential.
We give a brief sketch of this as follows and refer to \cite{bren,c3} for more details.  Thus one
wants to preserve $(\gD x)(\gD p)\geq \hbar^2/4$ for $x\sim x_k,\,\,p\sim p_k\,\,(k=1,2,3)$
and is led to the following transformation ($\ga\in{\bf R}$)
\bq\label{3.2}
(\gD x')^2=e^{-\ga}(\gD x)^2;\,\,(\gD p')^2=e^{-\ga}(\gD p)^2+\frac{\hbar^2}{4}
\frac{(e^{\ga}-e^{-\ga})}{(\gD x)^2}
\end{equation}
Consequently
\bq\label{3.3}
(\gD x')^2(\gD p')^2=e^{-2\ga}(\gD x)^2(\gD p)^2+\frac{\hbar^2}{4}(1-e^{-2\ga})
\end{equation}
Thus if $(\gD x)^2(\gD p)^2=\hbar^2/4$ it remains so for any $\ga$ and for $\ga\to\infty$
one has $(\gD x')^2(\gD p')^2\to\hbar^2/4$ for any value of $(\gD x)^2(\gD p)^2\geq(\hbar^2/4)$.
Now one considers a probability density P and an action variable S with functionals of the form
$({\bf 3B})\,\,{\mf A}=\int d^3xF(x,P,\na P,...,S,\na S,...)$ where classically $({\bf 3C})\,\,\pp_t{\mf A}=
\{{\mf A},H_C\}$ with $({\bf 3D})\,\,H_C=\int d^3x[P|\na S|^2/2m]$ a Hamiltonian functional and
\bq\label{3.4}
\{{\mf A},{\mf B}\}=\int d^3x\left[\frac{\gd{\mf A}}{\gd P(x)}\frac{\gd{\mf B}}{\gd S(x)}-
\frac{\gd{\mf B}}{\gd P(x)}\frac{\gd{\mf A}}{\gd S(x)}\right]
\end{equation}
This provides an infinite Lie algebra structure for functionals ({\bf 3B}).  The time transformations
are generated by $H_C$ applied to $P(x)$ and $S(x)$ and yields the continuity equation and the
HJ equation
\bq\label{3.5}
\pp_tP=-\na\cdot\left(\frac{P\na S}{m}\right);\,\,\pp_tS=-\frac{|\na S|^2}{2m}
\end{equation}
where $\na S=p$ is the classical momentum.
Now consider space dilatations $x\to exp(-\ga/2)x$  with
\bq\label{3.6}
P'(x)=e^{3\ga/2}P(e^{\ga/2}x);\,\,S'(x)=e^{-\ga}S(e^{\ga/2}x)
\end{equation}
noting that they keep the dynamical equations (3.5) invariant.  For simplicity assume that the
average momentum of the particle is zero; general results can then be retrieved by a
Galilean transformation.  Then the classical uncertainty for a momentum component is
$({\bf 3E})\,\,\gD p_{cl,k}^2=\int d^3x P(\pp_kS)^2$ and, dropping the index $k$, via (3.6)
$\gD p^2_{cl}$ transforms as $({\bf 3F})\,\,\gD' (p')^2_{cl}=e^{-\ga}\gD p^2_{cl}$
while $({\bf 3G})\,\,\gD (x')^2=e^{-\ga}\gD x^2$ (with $\gD x^2$ still unspecified).  Evidently
({\bf 3F}) shows that (3.2) does not hold but rather corresponds to the first term on the right
in (3.2).  Hence one must modify ({\bf 3E}) in order to get a quantity $\gD p^2$ satisfying
(3.2).  This leads to 
\bq\label{3.7}
\gD p^2_{q,k}=\int d^3x P(x)(\pp_kS(x))^2+\hbar^2{\mf Q}_k\,\,(k=1,2,3)
\end{equation}
Now impose the condition that the rules ({\bf 3H}) should transform $\gD p_q^2$ as prescribed
by (3.2) and this will reduce the set of possible functional forms of ${\mf Q}$.  There results
(cf. \cite{bren} for details)
\bq\label{3.8}
\gD (p'_q)^2=e^{-\ga}\gD p^2_{cl}+\hbar^2{\mf Q}'\Rightarrow \gD (p'_q)^2=e^{-\ga}\gD p^2_q
+\hbar^2({\mf Q}'-e^{-\ga}{\mf Q})
\end{equation}
Identifying this with (3.2) yields then
\bq\label{3.9}
{\mf Q}'-e^{-\ga}{\mf Q}=\frac{1}{4\gD x^2}(e^{\ga}-e^{-\ga})\Rightarrow {\mf Q}'-\frac{1}{4\gD (x')^2}
=e^{-\ga}\left({\mf Q}-\frac{1}{4\gD x^2}\right)
\end{equation}
The form of this equation indicates the existence of a relation between ${\mf Q}$ and $\gD x^2$
that is scale independent, namely $({\bf 3I})\,\,{\mf Q}_k=1/4\gD x_k^2$; this is the only possibility
for which the relation between $\gD p_k^2$ and $\gD x_k^2$ is independent of $\ga$.
In conclusion the supplementary term necessary to obtain a definition of $\gD p_q^2$
compatible with (3.2) is inversely proportional to $\gD x^2$ as in ({\bf 3I}).  Compatibility
with the Hall-Reginatto methods and techniques is then explained (cf. \cite{bren}) and 
one is led to the form 
\bq\label{3.10}
{\mf Q}_k=\gb\int d^3x[\pp_kP(x)^{1/2}]^2
\end{equation} 
leading to (for $\gb=1$)
\bq\label{3.11}
H_q=\int d^3x\left[\frac{P(x)|\na S(x)|^2}{2m}+\frac{\hbar^2}{2m}|\na P^{1/2}(x)|^2\right]
\end{equation}
and one has for $P=R^2$ (from $\psi=Rexp(iS/\hbar)$) the formula $\na P^{1/2}=
(1/2)\na P/P\Rightarrow |\na P^{1/2}|^2=(1/4)|\na P|^2$.  Hence the last term in (3.11)
coincides with a quantum potential times P via
\bq\label{3.12}
\frac{\hbar^2}{2m}(\na P^{1/2})^2=\frac{\hbar^2}{8m}\left(\frac{(\na P)^2}{P}\right);\,\,
Q=-\frac{\hbar^2}{2m}\frac{\gD P^{1/2}}{P^{1/2}};\,\,
\end{equation}
$$P Q=-\frac{\hbar^2}{8m}\left[2\gD P -\frac{(\na P)^2}{P}\right];\,\,\int PQd^3x=\frac{\hbar^2}{8m}
\int d^4x\frac{(\na P)^2}{P}$$
and this is the desired quantum addition to the classical Hamiltonian.
\\[3mm]\indent
{\bf REMARK 3.2.}
We note that in work of Gr\"ossing (cf. \cite{c7,gros}) one deals with subquantum thermal
oscillations leading to momentum fluctuations $({\bf 3J})\,\,\gd p=-(\hbar/2)(\na P/P)$
where P is a position probability density with $-\na log(P)=\gb \na{\mc Q}$ for ${\mc Q}$ a
thermal term (thus $P=cexp(-\gb {\mc Q})$ where $\gb=1/kT$ with $k$ the Boltzman constant).
This leads also to consideration of a diffusion process with osmotic velocity ${\bf u}\propto
-\na{\mc Q}$ and produces a quantum potential 
\bq\label{3.13}
Q=\frac{\hbar^2}{4m}\left[\na^2\tl{{\mc Q}}-\frac{1}{D}\pp_t\tl{{\mc Q}}\right]
\end{equation}
where $\tl{{\mc Q}}={\mc Q}/kT$ and $D=\hbar/2m$ is a diffusion coefficient.  Consequently
(cf. \cite{c7} one has a Fisher information $({\bf 3K})\,\,F\propto \gb^2\int exp(-\gb {\mc Q}
(\na {\mc Q})^2d^3x$.  As in the preceeding discussions the fluctuations are generated by the
position probability density and one expects a connection to (Bohmian) quantum mechanics
(cf. \cite{c3,crow,garb,gabr}).
$\hfill\bs$
\\[3mm]\indent
{\bf REMARK 3.3.}
There is considerable literature devoted to the emergence of quantum mechanics from
classical mechanics.  There have been many studies of 
stochastic and hydrodynamic models, or fractal situations, involving such situations and we
mention in particular \cite{c3,carr,c6,c7,crnt,crow,davi,garb,gabr,gran,gros,hall,hkr,hlal,kani,
kasc,nasi,nsti,nels,nttl,ntcl,olav,regi,schh,soni,tskv};
a survey of some of this appears
in \cite{c3}.  For various geometrical considerations related to the emergence 
question see also \cite{bfga,cgmo,cgml,elze,hily,hoft,issc,isio,isid,isgn,isdo,mssv}.
$\hfill\bs$

\end{document}